\documentstyle[pra,manuscript,aps]{revtex}
\draft
\tighten
\begin{document} 
\title{Theory of Ferromagnetic Metal to Paramagnetic Insulator Transition
in $R_{1-x}A_x$MnO$_3$} 
\author{L. Sheng$^1$, D. N. Sheng$^{1,2}$ and C.S. Ting$^{1,2}$} 
\address{$^1$Physics Department and Texas Center for Superconductivity, 
University of Houston, Houston, TX 77204\\
$^2$National Center for Theoretical Sciences P.O. Box 2-131,
Hsinchu, Taiwan 300, R.O.C} 
\maketitle 
\begin{abstract} 
The double-exchange model for the Mn oxides
with orbital degeneracy
is studied with including on-site Coulomb repulsion,
Jahn-Teller (J-T) coupling and doping-induced
disorder. In the strong interaction limit, it is mapped onto a 
single-band Anderson model, in which all scattering
mechanisms can be treated on an equal footing.
A sharp rise in the mean square fluctuation  
of lattice distortions is found near the Curie temperature
$T_c$, in agreement
with experiments. We show that the spin and J-T disorders
lead to a metal-insulator transition (MIT)
only at low carrier density. The 
MIT observed in samples with $0.2\leq x<0.5$
can be explained by further including the disorder
effect of cation size mismatch. 
\end{abstract} 
\mbox{}\\ 

The importance of electron-lattice coupling 
to the transport properties of the 
Mn oxides $R_{1-x}A_x$MnO$_3$, which exhibit
colossal magnetoresistance (CMR) effect~\cite{s1,s2,s3,s4}
in the hole-doping range $0.2\le x< 
0.5$, was pointed out in early theoretical works~\cite{s5,s6}.
Experimental measurements~\cite{s7,s8,s9,s10} have 
repeatedly given evidences for the existence of 
Jahn-Teller (J-T) effect in the Mn oxides. 
In those works, the novel ferromagnetic metal (where $d\rho/dT>0$
with $\rho$ as the resistivity) to paramagnetic insulator
($d\rho/dT<0$) transition was understood
as a crossover from large to self-trapped
 small polarons caused by the
band narrowing accompanying  
with local spin disordering during the transition. 
Such a description does make sense if the carrier
density is low enough so that the small polarons
are well separated in space. 
In the high density regime, e.g., for the Mn oxides with
intermediate doping, there should be large overlaps
among the polaron wave functions, and 
the validity of small polaron picture becomes questionable. 
On the other side, recent experiments~\cite{s10} showed that large 
and small polarons actually coexist in the insulating 
paramagnetic phase, an indication that even large polarons
can also be localized and do not contribute to metallic
conductivity. These issues suggest that a more accurate
criterion for the metal-insulator transition (MIT)
is urgently needed.

A valid theory for the MIT at
intermediate doping has to be based upon a
unified nonperturbative treatment of these equally important
scattering mechanisms: (i) strong double-exchange (DE)
interaction between conduction $e_g$ electrons and localized
spins~\cite{s11,s12,s13,s14}, 
(ii) J-T coupling and on-site 
Coulomb repulsion associated with doubly degenerate
$e_g$ bands, and (iii) doping-induced nonmagnetic
disorder. Factor (iii) playing an 
important role in the transport properties
of the Mn oxides has been well demonstrated in
experiments~\cite{s3,s4}. It was proposed there that  
the size mismatch between $R$ and $A$ atoms
leads to bending of local Mn-O-Mn bonds, and hence strongly
reduces the electron hoping integral.
With the basic structure of the MnO$_6$ octahedra
being similar, the compounds $R_{1-x}A_x$MnO$_3$ with different
$R$ and $A$ should have about the same
J-T coupling strength. Therefore, tunning the cation size
mismatch has become a practical way to control their 
transport properties~\cite{s4}.  
Previous theoretical works on the MIT focused on only one or 
two of these features. For example, in Refs.\ [5] and [6] the lattice
effect has been emphasized but the on-site Coulomb interaction
and doping-induced disorder were neglected. These theories
failed to account correctly for the doping dependence of the 
MIT. On the other hand, nonmagnetic
disorder effect was considered within the 
single-band DE model in Refs.\ [15] and [16]
with the J-T distortions being omitted.  The works [15] and [16]
overestimated the strength of nonmagnetic
disorder necessary for the occurrence of MIT. 
A complete theory, which is able to describe the 
essential physics of the MIT in the Mn oxides,
is still lacking. 

In the present work, the
MIT in the Mn oxides will be investigated
by incorporating all scattering
mechanisms outlined above. Employing the path-integral 
approach, we map our Hamiltonian   
onto a single-band Anderson model
in the strong interaction limit.  From this model, the MIT  
can be numerically studied using the transfer matrix method
without further approximation~\cite{s17}. The J-T coupling
strength is estimated
by comparing our calculated resistivity with the experimental
data of La$_{1-x}$Sr$_{x}$MnO$_3$. For the first time,
the role of the J-T distortion on the MIT has been determined 
very precisely. The phase diagram obtained
describes consistently the behavior of the MIT
observed in samples with CMR as a function of doping
concentration and cation size mismatch.  

In the Mn oxides, each Mn atom has five outer-shell
$3d$ orbitals, three half-filled $t_{2g}$ states giving
rise to an $S=3/2$ localized spin, and two $e_g$ states
$|+\rangle=|x^2-y^2\rangle$ and $|-\rangle=|
3z^2-r^2\rangle$ forming two-fold degenerate conduction
bands. Strong Hund's rule coupling forces electron spins
on a site to polarize completely, so we can write down a
Hamiltonian, in which the conduction electrons are
effectively spinless,
\begin{eqnarray}
H=-\sum\limits_{ij}f_{ij}
({\bf d}_{i}^\dagger\hat{t}_{ij}{\bf d}_j)
+\sum\limits_i U'n_{i+}n_{i-} 
-g\sum\limits_i({\bf d}_i^\dagger
\mbox{\boldmath{$\tau$}}{\bf d}_i)\cdot{\bf Q}_i+\frac{k}{2}
\sum\limits_i Q_i^2\ .
\end{eqnarray}
The first term represents the two-band DE model~\cite{s13,s16}, 
where ${\bf d}_i^\dagger=(d_{i+}^\dagger,d_{i-}^\dagger)$
and $f_{ij}=\cos(\theta_i/2)
\cos(\theta_j/2)+\sin(\theta_i/2)\sin(\theta_j/2)
e^{-i(\varphi_i-\varphi_j)}.$ 
The second term stands for the on-site
inter-orbital Coulomb repulsion. 
The third and fourth terms describe the coupling between
electrons and two J-T distortion modes 
${\bf Q}_i=Q_x\hat{\bf i}
+Q_z\hat{\bf k}$, and the harmonic lattice
deformation energy, respectively~\cite{s5}, where
$\mbox{\boldmath{$\tau$}}$ are the Pauli matrices.
The hopping integral matrix $\hat{t}_{ij}$ can be 
written as~\cite{s18}
\begin{equation}
\hat{t}_{ij}=(t-r_{ij}\Delta t)(1+\mbox{\boldmath{$\tau$}}
\cdot{\bf n}_{ij})\ ,
\end{equation}
where $t$ is the bare hopping integral, 
$r_{ij}\Delta t$ is used to model    
the disorder originated from cation size
mismatch with $\Delta t$ as
its amplitude and $r_{ij}$ random numbers between $0$ and $1$.
Since $R$ and $A$ atoms are randomly distributed,
their size mismatch not only decreases the
average electron band width but also leads to off-diagonal disorder.
Here, ${\bf n}_{ij}={\bf n}_\alpha$ with $\alpha=x,y$ and $z$
for hopping along the $x$, $y$ and $z$ directions, and ${\bf n}_x
=-(\sqrt{3}\hat{\bf i}+\hat{\bf k})/2$, ${\bf n}_y=
(\sqrt{3}\hat{\bf i}-\hat{\bf k})/2$, ${\bf n}_{z}=\hat{\bf k}$. 

Since both the J-T coupling and on-site Coulomb interaction
disfavor double occupancy, they have the common tendency to 
split the degenerate $e_g$ bands. From Eq.\ (1), 
it is easy to derive a
mean-field Stoner criterion for the occurrence of 
band splitting at half-filling $(x=0)$
\begin{equation}
(U'+2E_J)D(0)>1\ ,
\end{equation}
where $E_J=g^2/k$, and $D(0)$ is the single-band
density of states at the band center before the 
band splitting.  Equation (3) is also the criterion
at which static J-T distortions $(\langle{\bf Q}_i\rangle\neq 0)$
occurs. In the absence of Coulomb repulsion, it needs  
relatively strong J-T coupling to split the $e_g$ 
bands and induce static J-T distortions.
In the opposite limit where 
$U'$ is larger than the band width, 
any small J-T coupling will lead to static
J-T distortions. The on-site Coulomb repulsion
enhances substantially the J-T effect.
This is consistent with the argument of 
Varma that the Mn oxides at  
half-filling $(x=0)$ are Mott insulators, while
the J-T effect occurs parasitically~\cite{s14}. 

According to experiments, 
the condition in Eq.\ (3) should be well
satisfied in the Mn oxides. In order to ensure
the system being insulator at $x=0$ and to reduce
the number of involved parameters,
we shall project out double occupancy by setting $(U'+2E_J)
\rightarrow\infty$. In the path-integral representation,
the isospin-charge separation transformation 
${\bf d}_i=c_i{\bf y}_i$ can be introduced, 
where ${\bf y}_i$ is the $CP^1$ field for the isospin 
and $c_i$ is the charge operator~\cite{s19}.  
The effective Lagrange corresponding to Hamiltonian
Eq.\ (1) is then derived to be
\begin{eqnarray}
L(\tau)&=&\sum\limits_i\Bigl[c_i^\dagger c_i{\bf y}_i^\dagger
\partial_\tau{\bf y}_i+c_i^\dagger(\partial_\tau-\mu)c_i\nonumber\\
&-&\sum\limits_{j}f_{ij}
({\bf y}_i^\dagger\hat{t}_{ij}
{\bf y}_j)c_i^\dagger c_j
-g{\bf o}_i\cdot{\bf Q}_i
c_i^\dagger c_i+\frac{k}{2} Q_i^2\Bigr]\ ,
\end{eqnarray} 
where ${\bf o}_i(\tau)={\bf y}_i^\dagger(\tau)
\mbox{\boldmath{$\tau$}}{\bf y}_i(\tau)$ is a unit vector 
characterizing the instantaneous isospin orientation. 
The third term in Eq.\ (4) indicates
that the isospin kinetic motion is characterized by the
energy scale $K_0=t\langle c_i^\dagger c_j\rangle$,
which for $0.2\leq x<0.5$ is numerically obtained as  
$K_0\simeq 0.10t\sim 0.16t$ at $\Delta t=0$. 
Disorder will further decrease this value. 
On the other hand, the J-T coupling given by the fourth
term in Eq.\ (4) is estimated to be  
$-E_J$.  Therefore, as long as $E_J>t$, the J-T coupling
dominates and parallel configuration
between ${\bf Q}_i$ and ${\bf o}_i$ is energetically 
favorable; namely, ${\bf o}_i={\bf Q}_i/Q_i$ being
confined to the $x-z$ plane.
A permitted parametrization~\cite{s19} 
of ${\bf y}_i^\dagger$ is ${\bf y}^\dagger_i=[\cos(\phi_i/2),
\sin(\phi_i/2)]$ with $-\pi<\phi_i\leq\pi$, where 
$\phi_i$ is the angle between the isospin and the 
$z-$axis.

To study the statistics of the J-T distortions,
we can carry out the Gaussian integral of the 
charge variables $c_i(\tau)$ in Eq.\ (4). The saddle point
equation of the resulting Lagrange yields  
$\langle Q_i\rangle=g(1-x)/k$. Then expanding 
the Lagrange around $\langle Q_i\rangle$ up to 
quadratic order in 
$\delta Q_i=Q_i-\langle Q_i\rangle$, we obtain a Gaussian 
distribution $P(\varepsilon_i)=\exp(-\varepsilon_i^2
/2\Delta_J^2)/\sqrt{2\pi}\Delta_J$ for the dynamical
fluctuations of J-T distortions with 
$\varepsilon_i=-g\delta Q_i$, where 
\begin{equation}
\Delta_J^2=(k_BT)E_J/[1-E_JD(\mu)]
\end{equation}
is equal to  $\langle \varepsilon^2_i\rangle=g^2
\langle(\delta Q_i)^2\rangle$. For fixed $E_J$,
the amplitude $\Delta_J/g$ of lattice fluctuations
depends on temperature $k_BT$ and electron
density of states $D(\mu)$ at the chemical potential $\mu$ for 
$\delta Q_i\equiv 0$. The denominator of Eq.\ (5)
represents the dynamical aspect of the lattice
distortions due to electron density fluctuations within the 
random phase approximation (RPA). According to Eq.\ (4), 
we obtain an effective single-band Hamiltonian  
\begin{equation}
H_{\mbox{\small eff}}
=-\sum\limits_{ij}\widetilde{t}_{ij}c_i^\dagger c_j
+\sum\limits_i\varepsilon_ic_i^\dagger c_i\ ,
\end{equation}
where $\widetilde{t}_{ij}=f_{ij}
({\bf y}_i^\dagger\hat{t}_{ij}{\bf y}_j)$, and 
a constant energy $-(1-x)^2E_J/2$ per site has been omitted.
With explicit parametrization, $\widetilde{t}_{ij}$ can 
be written as
\begin{eqnarray}
\widetilde{t}_{ij}=f_{ij}(t-r_{ij}\Delta t)
\bigl\{\cos[(\phi_i-\phi_j)/2]
+\cos[(\phi_i+\phi_j)/2-\gamma_\alpha]\bigr\}\ ,
\end{eqnarray}
with $\gamma_\alpha=2\pi/3$, $-2\pi/3$ and $0$ for electron
hopping in the $x$,  $y$ and $z$ directions. Equation (7)
is the effective electron hopping integral renormalized
by spin disorder, ionic size mismatch and orbital disorder. 

To investigate the magnetic and orbital ordering transitions,
we replace $c_i^\dagger c_j$ in Eq.\ (6)
by its average $K_0/t$. The resulting Hamiltonian for the 
spins and isospins is studied by Monte Carlo simulation
on a $20\times 20\times 20$ site lattice.
We find the spin and isospin ordering transition temperatures
$T_c$ and $T^*$, respectively, to be
\begin{equation}
T_c\simeq 1.3K_0\ ,\mbox{}\hspace{8mm}
T^*\simeq 0.5T_c\ .
\end{equation}
If the bare band width $W=12t$ is taken to be $2eV$,
then $T_c$ is about $250K$ to $400K$ for $0.2\le x<0.5$ 
in the absence of disorder.  
The disorder effect 
may still lower $T_c$ by a finite amount~\cite{s16}. The value of 
$T_c$ evaluated here is comparable to experimental data for the
Mn oxides. Our finite-size calculation seems to indicate that 
the isospins are ordered below $T^*$, while the result
based on a mean-field
approximation implies only strong short-range ordering~\cite{s18}.
 
The Hamiltonian Eq.\ (6) represents an Anderson model
with temperature-dependent diagonal and off-diagonal
disorders. Well below $T_c$, the spins are ordered
and the fluctuations of the on-site J-T energy $\varepsilon_i$
are relatively weak, so we expect a metallic phase
there. As the temperature is increased to above $T_c$,
the local spins rapidly become disordered, leading to 
DE off-diagonal disorder.  Besides, along with
the local spin disordering, 
the electron band narrows and $D(\mu)$ increases.
From Eq.\ (5), it follows that the fluctuation amplitude
$\Delta_J$ of $\varepsilon_i$ also increases near $T_c$.
For simplicity, we 
assume the local spins $(\theta_i,\varphi_i)$
to be completely ordered and disordered
below and above $T_c$, respectively.
The isospins $(\phi_i)$ are disordered  
above $T^*=0.5T_c$ according to Eq.\ (8). 
If $E_J$ is taken to be $E_J=3.7t$ (see later discussion), 
by exact diagonalization of Eq.\ (6) with $\varepsilon_i
\equiv 0$ on a lattice with $10\times 10\times 10$
sites to calculate $D(\mu)$, it is found that $\Delta_J$
increases by $30\%$ at $\Delta t=0$
during the magnetic transition.
This result is consistent with the extensive
experimental observations~\cite{s7,s8,s9}
of rapidly enhanced lattice fluctuations near $T_c$.
The increased spin and lattice disorders 
may possibly lead to an Anderson MIT. The picture of
Anderson localization is applicable to finite temperatures,
as long as the electron localization lengths are smaller
than the dephasing length $\ell_{\mbox{\small in}}$
due to inelastic scattering.
This condition could be expected in the Mn oxides
in the experimental temperature range; otherwise
their insulator transport behavior could not be observed.

Before determining the localization
effect due to Eq.\ (6), we need to estimate
the J-T coupling $E_J(=k^2/g)$ by comparing our calculated
resistivity with experimental data. Since 
the effect of DE spin disorder and J-T distortions 
is most prominent in the cleanest
systems, we consider La$_{1-x}$Sr$_{x}$MnO$_3$~\cite{s9},
where the ionic mismatch
can be assumed negligible ($\Delta t\simeq 0$). 
According to the sign of $d\rho/dT$, one finds the paramagnetic
phase of La$_{1-x}$Sr$_{x}$MnO$_3$ becomes obviously metallic
in the doping range $x=0.3\sim 0.4$ with 
resistivity of the order of $10^{-3}\Omega$cm 
at $T=400K$~\cite{s9}.  On the other hand, we 
can calculate the resistivity from Eq.\ (6) numerically.
By using the well-known transfer matrix method
developed by Mackinnon and Krammer~\cite{s17} 
to calculate the mobility edge,
we first determine the phase diagram in the $n_e(=1-x)$
vs $\Delta_J$ plane at $\Delta t=0$ in the
paramagnetic phase, as given in Fig.\ 1(a).
Here, the disorder coming from both the randomly orienated
localized spins and J-T distortions has been
considered.  In the corresponding metallic 
region, we then apply the Landauer formula~\cite{s20}
to calculate the conductivity 
at $T>T_c$ as a function of $x$ for different values of
$\Delta_J$, as plotted in Fig.\ 1(b),
where the experimental data at $T=400K$ of Ref.\ [9] are shown
by squares. For $0.2\le x<0.5$, it is found that $D(\mu)$
changes very slowly with electron density
$n_e$, so for a given temperature
$\Delta_J$ is approximately constant.
By comparison of the calculated and experimental data in 
Fig.\ 2(b), we find $\Delta_J{\simeq}
1.5t$ at $T=400K$ for the Mn oxides, which means $E_J\simeq 3.7t$.
This value of $E_J$ corresponds 
to $\lambda=1.1$ in the theory of Millis 
$et$ $al.$~\cite{s5}. According to Fig. 1(a), 
for such strength of J-T coupling, the DE spin disorder
and J-T distortions lead to an insulator paramagnetic
phase only for $n_e\geq 0.88$ or $x\leq 0.12$.
It is important to notice that since no additional
scattering mechanisms besides spin disorder and J-T
distortions are  considered here, the estimated J-T
coupling strength $E_J=3.7t$ should be regarded as an
upper bound.

In order to study the MIT for samples with $x>0.12$,
the effect of the cation size mismatch or nonzero
$\Delta t$ must be included. 
The electron density of states and mobility edge
are calculated numerically by finite-size 
diagonalization and scaling calculation~\cite{ss1}.
From the calculated results, we obtain the phase diagram Fig.\ 2 
in $n_e$ vs $\Delta t$ plane for the MIT. 
In the limit of large $\Delta t$,  
the denominator of Eq.\ (5) may possibly go to 
zero or negative, giving rise to the 
RPA instability. Within mean-field approximation,
it is easy to see that the RPA instability simply
means $\delta \mu/\delta n_e\leq 0$, an indication of 
phase separation~\cite{s21} into 
hole-rich regions with relatively weak lattice distortions
and hole-poor regions with strong lattice distortions.
Such unstable region is also shown in Fig.\ 2.
It is clear that the MIT occurs
in broad ranges of values of $n_e$ and $\Delta t$.
With increasing $\Delta t$, localized 
hole number increases and Anderson transition always
precedes phase separation. The phase diagram Fig.\ 2
is in agreement with the experimental observation~\cite{s3,s4}
that the transport properties of the Mn oxides could be drastically
altered by tunning the ionic size mismatch. 
To further determine the system properties in the phase-separation regime,
one needs to consider anharmonic deformation energy of the lattice
and take nonuniform states into account. The residue resistivity coming
from the cation size mismatch in 
the low-temperature metallic state can also be calculated.
The critical residual 
resistivity as shown by the open circles in Fig. 2
for occurrence of a MIT is found to be the order of
$10^{-4}\Omega$cm or less, being smaller than that
estimated in the absence of J-T effect~\cite{s16}. 
The present result seems to be more reasonable, and is 
comparable to data from epitaxial films~\cite{s22} or
single crystal systems~\cite{s9}.

In summary, employing the path-integral approach and 
numerical scaling calculations, we have shown that  
the essential physics of the 
MIT in the Mn oxides with $0.2\leq x<0.5$ 
can be understood as an Anderson transition driven by DE
spin disorder, J-T distortions and cation size
mismatch. The phase separation
predicted in Fig. 2 for strong cation size mismatch is also
consistent with  the inhomogeneous electron
states observed recently near and above $T_c$ in
the certain Mn oxides~\cite{s23}.

This work was supported by the Texas Center for Superconductivity
at the University of Houston, by the Robert A. Welch foundation,
and by the National Research Council in Taiwan.

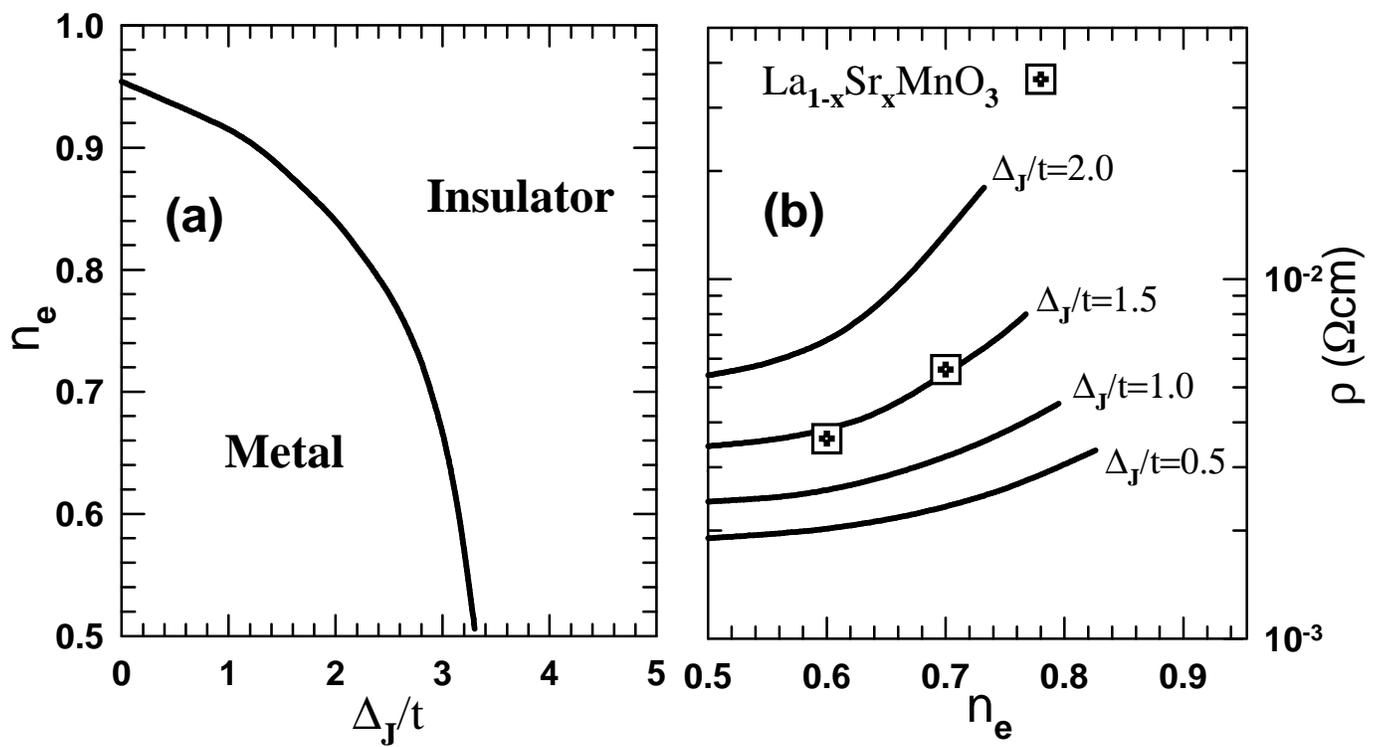
\begin{figure}
\caption{(a) Phase diagram in $n_e$ vs $\Delta_J/t$ plane
in the paramagnetic phase where $n_e=1-x$ is the electron 
density, and (b) the resistivity in the metallic paramagnetic
phase as a function of $n_e$ for several values of 
$\Delta_J/t$ at $T=400K$.}
\end{figure}
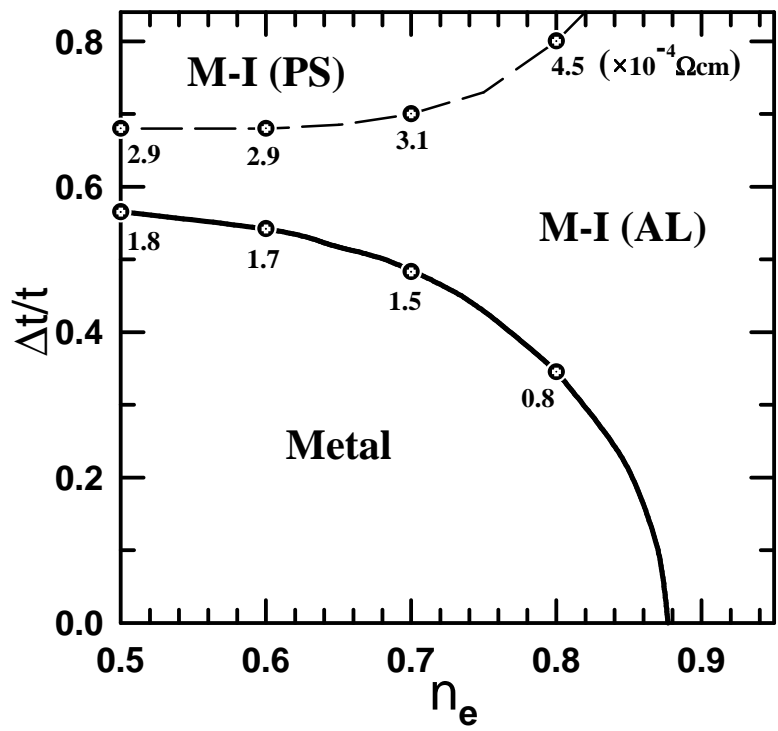
\begin{figure}
\caption{Phase diagram for the metal-insulator transition in  
$\Delta t/t$ vs $n_e$ plane for $E_J=3.7t$.
In the region labeled ``Metal''
the system is metallic in both the ferromagnetic
and paramagnetic phases, while in the regions labeled ``M-I(AL)''
and ``M-I(PS)'' the system undergoes a metal-insulator
transition near $T_c$, where AL or PS indicates Anderson
localization or phase separation as the origin of
the insulator paramagnetic phase. The corresponding
critical residual resistivity at some points indicated by
small circles is shown in unit $10^{-4}\Omega$cm.} 
\end{figure}
\end{document}